\documentclass[twocolumn,showpacs,preprintnumbers,amsmath,amssymb,aps]{revtex4-1}
\usepackage{graphicx}
\usepackage{dcolumn}
\usepackage{bm}

\begin{document}

\title{A plausible method of preparing the ideal {\it p-n} junction interface of a thermoelectric material by surface doping}

\author{Ji-Eun Lee$^{1,2}$}
\author{Jinwoong Hwang$^{1,3}$}
\author{Minhee Kang$^{1}$}
\author{Hyun-Jeong Joo$^{1}$}
\author{Hyejin Ryu$^{2}$}
\author{Kyoo Kim$^{4,5}$}
\author{Yongsam Kim$^{6}$}
\author{Namdong Kim$^{6}$}
\author{Anh Tuan Duong$^{7}$}
\author{Sunglae Cho$^{7}$}
\author{Sung-Kwan Mo$^{3}$}
\author{Choongyu Hwang$^{1}$}\email{ckhwang@pusan.ac.kr}
\author{Imjeong Ho-Soon Yang$^{1}$}\email{hsyang@pusan.ac.kr}

\affiliation{$^1$ Department of Physics, Pusan National University, Busan 46241, South Korea}
\affiliation{$^2$ Center for Spintronics, Korea Institute of Science and Technology, Seoul 02792, South Korea}
\affiliation{$^3$ Advanced Light Source, Lawrence Berkeley National Laboratory Berkeley, California 94720, USA}
\affiliation{$^4$ Korea Atomic Energy Research Institute, 111, Daedeok-Daero 989Beon-Gil, Yuseong-Gu, Daejeon 34057, South Korea}
\affiliation{$^5$ Max Planck-POSTECH/Hsinchu Center for Complex Phase Materials. Max Plank POSTECH/Korea Research Initiative (MPK), Gyeongbuk 37673, South Korea}
\affiliation{$^6$ Pohang Accelerator Laboratory, Pohang University of Science and Technology, Gyeongbuk 37673, South Korea}
\affiliation{$^7$ Department of Physics, University of Ulsan, Ulsan 44610, South Korea}

\begin{abstract}
Recent advances in two-dimensional (2D) crystals make it possible to realize an ideal interface structure that is required for device applications. Specifically, a {\it p-n} junction made of 2D crystals is predicted to exhibit an atomically well-defined interface that will lead to high device performance. Using angle-resolved photoemission spectroscopy, a simple surface treatment was shown to allow the possible formation of such an interface. Ta adsorption on the surface of a {\it p}-doped SnSe shifts the valence band maximum towards higher binding energy due to the charge transfer from Ta to SnSe that is highly localized at the surface due to the layered structure of SnSe. As a result, the charge carriers of the surface are changed from holes of its bulk characteristics to electrons, while the bulk remains as a {\it p}-type semiconductor. This observation suggests that the well-defined interface of a {\it p-n} junction with an atomically thin {\it n}-region is formed between Ta-adsorbed surface and bulk.
\end{abstract}

\maketitle

\section{Introduction}

Emergence of new materials invites their application to conventional devices with enhanced functionality~\cite{novoselov}. Indeed, successful preparation of two-dimensional (2D) crystals has led to the fabrication of a 2D materials-based {\it p-n} junction, which provides high device efficiency due to higher charge carrier mobility and cleaner interface, compared to those of the junction based on three-dimensional (3D) materials~\cite{henck,eschbach,coy,lee,si,frisenda,Pierucci, ZhangACS, ZhangNanoscal}. However, the device fabrication methods including artificial stacking, external gate, and chemical vapor deposition (CVD)~\cite{frisenda,dean,chavez,basko} still restrain device functionality due to lattice mismatch, insufficient modulation of charge carriers, defects that are trapped at the interface, etc~\cite{lin,bruno}. 

One of plausible methodologies to overcome these obstacles could be to induce a sharp gradient of the charge carrier type within a single material. SnSe could be a good candidate for this purpose. SnSe is a van der Waals material with negligible overlap between the wavefunctions of a charge carrier from adjacent layers, hence providing completely different chemical environments from the covalently bonded interface, corresponding to the 3D materials-based junctions. In addition, SnSe provides an excellent controllability of its charge carrier type, based on the recent effort to realize high performance thermoelectric devices~\cite{chang,zhaoNature,zhaoScience}.

The type of charge carriers of SnSe can be controlled by introducing impurities. The formation of Sn vacancies~\cite{jaekwang} and the substitution of Sn with Na~\cite{maeda,lu} result in hole doping, while Br or Bi substitution~\cite{chang,duong} brings about electron doping, all of which are determined in the process of sample growth. Alternatively, it can also be achieved after the sample growth, for example, by the adsorption of foreign atoms such as K~\cite{zhang}. A previous theoretical study also predicts that the formation of a metal-SnSe junction can induce {\it n}- or {\it p}-type charge carriers for Ag, Au, or Ta~\cite{xian}. 

Here, {\it n}-type doping of the surface of a {\it p}-type semiconducting SnSe single crystal has been examined by direct measurement of the electron band structure using angle-resolved photoemission spectroscopy (ARPES) in conjunction with first principles calculations. While bulk SnSe exhibits {\it p}-type charge carriers due to Sn vacancies~\cite{jaekwang}, Ta adsorption leads to the shift of the whole band structure towards higher binding energy that is as much as 0.2~eV only with 0.025~monolayer (ML) of Ta, converting the type of charge carriers from {\it p}-type that of the bulk to {\it n}-type, consistent with the theoretical results. Based on the comparison to the previous theoretical study~\cite{xian}, highly efficient and surface localized {\it n}-type doping has been realized by Ta adsorption, while the bulk remains as a {\it p}-type semiconductor. This finding provides a viable route towards the fabrication of an atomically well-defined and hence highly efficient vertical {\it p-n} junction interface via surface treatment.

\section{Methods}

\subsection{Experimental details}
The $p$-type SnSe samples were prepared using a temperature gradient growth method detailed elsewhere~\cite{kim}. High purity Sn and Se powders were sealed in a quart ampoule, followed by annealing at 930~K for 10~hours. Thus prepared SnSe single crystal has been cleaved in an ultra-high vacuum chamber for Ta adsorption and {\it in-situ} angle-resolved photoemission spectroscopy (ARPES) measurements. Ta was deposited on the cleaved surface of SnSe using an e-beam evaporator and its coverage was estimated using a quart crystal thickness monitor. ARPES measurements were performed at the beamline 10.0.1.1 of the Advanced Light Source in Lawrence Berkeley National Laboratory using a photon energy of 65~eV. Measurement temperature was 50~K, while energy and momentum resolutions were 40~meV and 0.01~${\rm \AA}^{-1}$, respectively. The X-ray diffraction (XRD) data were taken at BL1C of the PLS-II in Pohang Accelerator Laboratory. 

\subsection{Electron band structure calculations}
The density functional theory was adopted to obtain the electron band structure within Perdew–Burke–Ernzerhof exchange correlation functional. Tran Blaha-modified Becke Johnson potential (TB-mBJ) is used as an inexpensive remedy to the band gap underestimation problem of the DFT~\cite{Tran}. The band gap obtained in this way is consistent with previous calculations using TB-mBJ and {\it GW} methods~\cite{Hong,Shi}. The effect of Ta adsorption was taken into account by utilizing the Korringa-Kohn-Rostoker (KKR) green function method~\cite{Zeller} combined with coherent potential approximation method~\cite{Soven} implemented in the AkaiKKR code~\cite{Akai}. In the actual calculation, Ta atoms were introduced inside the van der Waals gap of bulk SnSe. 

\begin{figure}[b]
\centering
\includegraphics[width=1\columnwidth]{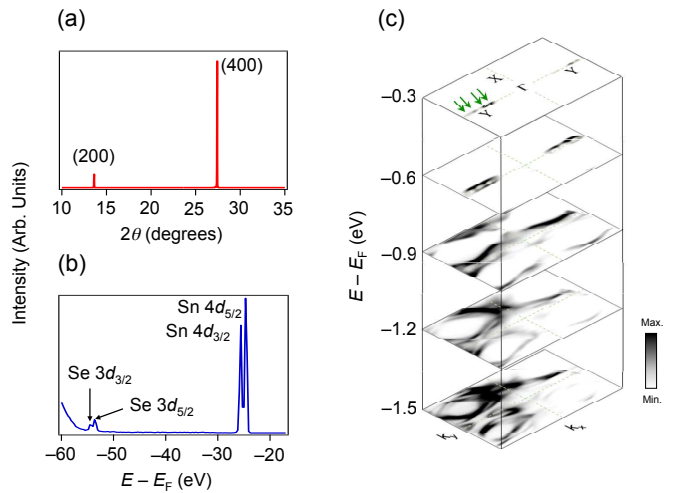}
\caption{\label{fig:fig1} (Color online) (a) X-ray diffraction data for single crystal SnSe with Sn-vacancy. (b) X-ray photoemission spectrum for single crystal SnSe with Sn-vacancy. (c) Constant energy maps taken using ARPES at $E-E_{\rm F} = -0.3, -0.6, -0.9, -1.2, {\rm and} -1.5~{\rm eV}$.}
\end{figure}

\section{Results and discussions}

Sample quality has been examined using XRD and x-ray photoemission spectroscopy (XPS). Figure~1(a) shows an XRD pattern of the SnSe sample that shows only ({\it h}00) peaks, which are (200) and (400), indicating single crystalline nature of the SnSe sample. The observed peaks provide a lattice constant along the {\it c}-axis of 11.655~{\rm \AA} from the Bragg's law, that is comparable to 11.483~{\rm \AA} reported in a previous work~\cite{duong}. Figure~1(b) shows Se 3$d$ and Sn 4$d$ states obtained in an ultra-high vacuum chamber after cleaving along the $c$-axis, along which layered SnSe is stacked by the weak van der Waals force. Each core-level spectrum shows spin-orbit doublet, without showing any other impurities within the range of the measurement. Figure~1(c) shows constant energy maps taken from the same sample at several different energies relative to the Fermi energy, $E_{\rm F}$, using ARPES. At $E-E_{\rm F}=-0.3~{\rm eV}$ that is close to the valence band maximum (VBM), four spots are observed along the $\Gamma-{\rm Y}$ direction as denoted by green arrows, indicating the multi-valley valence band, characteristics of SnSe~\cite{lu}. Upon decreasing energy, four spots grow in size, especially as shown in the constant energy map taken at $-0.6~{\rm eV}$, indicating the hole-like feature of the quasiparticles of SnSe near the VBM, consistent with previous ARPES results~\cite{maeda,lu,zhang,wang}.

\begin{figure*}
\centering
\includegraphics[width=2\columnwidth]{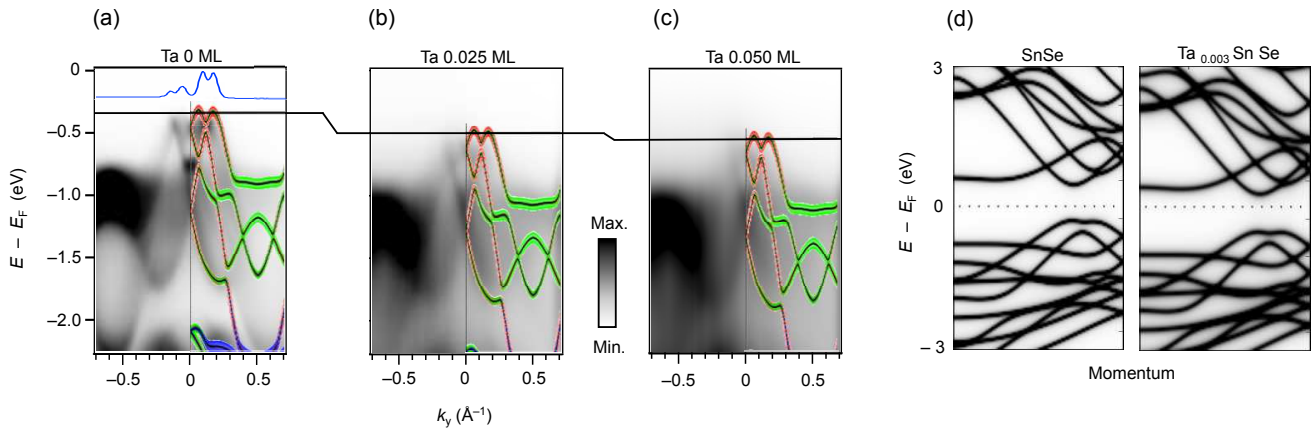}
\caption{\label{fig:fig2} (Color online) (a-c) ARPES intensity maps of SnSe taken along $\Gamma-{\rm Y}$ direction for three different Ta coverages, 0~ML, 0.025~ML, and 0.050~ML. The blue curve is a momentum distribution curve taken at the valence band maximum, i.\,e.\,, 0.32~eV below $E_{\rm F}$. The calculated electron band structure of SnSe by the TB-mBJ method is rigidly shifted to fit to the ARPES data. Red, green, and blue color in the calculated band denotes Se $p_{x}$, $p_{y}$, and $p_{z}$ orbitals, respectively. The black line denotes the shift of the valence band maximum. (d) The calculated electron band structure of SnSe with and without Ta by the KKR green function method.}
\end{figure*}

The effect of Ta adsorption on the electron band structure of SnSe is directly observed using ARPES. Figure~2(a) shows an ARPES intensity map taken from the pristine SnSe (Ta 0~ML) along $k_{\rm y}$-axis ($\Gamma-{\rm Y}-\Gamma$ direction) at $k_{\rm x}=0.5~{\rm \AA}^{-1}$, denoted in Fig.~1(c). The black curves are calculated band structure of SnSe using the TB-mBJ method. Red, green, and blue symbol denotes Se $p_{x}$, $p_{y}$, and $p_{z}$ orbitals, respectively. Although the photoelectron intensity near the VBM is very weak, the comparison between experimental and calculated results shows that the VBM of pristine SnSe is estimated to be 0.32~eV below $E_{\rm F}$ and it mostly exhibits an Se $p_{x}$ orbital character. When the previous optical measurements suggest that the energy gap of SnSe is 0.9~eV~\cite{wang,parenteau}, the VBM of undoped SnSe is supposed to be at $0.45~{\rm eV}$ below $E_{\rm F}$. As a result, the observed VBM at 0.32~eV below $E_{\rm F}$ suggests that the pristine SnSe is a {\it p}-type semiconductor, whose origin is Sn vacancies that take charges from the SnSe sample~\cite{jaekwang}. Different $E_{\rm F}-E_{\rm VBM}$ from a previous study of $0.2~{\rm eV}$ indicates that the sample used in the experiments shown in Fig.~2(a) exhibits less defects, resulting in less hole-doping with respect to the undoped sample. A momentum distribution curve taken at the VBM, the blue curve in panel (a), clearly shows 4 peaks, revealing the multi-valley valence band of SnSe~\cite{lu}. While the electron band structure close to the VBM exhibits 4 humps and 3 dips, that looks similar to the bow-shaped electron band structure observed in monolayer GaSe~\cite{Aziza}, a clear difference is that GaSe shows 2 humps and 1 dip at the $\Gamma$ point due to spin-orbit coupling whereas we verify that spin-orbit coupling barely changes the electron band structure close to the VBM~\cite{Nagayama}.

Ta adsorption leads to the shift of the whole band structure towards the higher binding energy. Especially, despite small amount of Ta atoms, e.\,g.\,, 0.025~ML, is adsorbed on the SnSe surface, the shift reaches as much as $\sim0.2~{\rm eV}$ as shown in Fig.~2(b). The band structure further shifts towards the higher binding energy with further Ta adsorption as shown in Fig.~2(c), typical of {\it n}-doping induced by metal adsorption. The ARPES intensity map becomes fuzzy upon introducing Ta, which might be due to the formation of random site potentials induced by Ta adsorption, that worsens the energy and momentum resolution of ARPES measurements. It is noteworthy that the observed shift is in contrast to the previous theoretical prediction for the Ta/SnSe interface~\cite{xian}. Instead, the observed overall shift is in agreement with our calculated results by the KKR green function method as shown in Fig.~2(d). The calculated electron band structure of SnSe shows $E_{\rm F}$ almost at the center of the indirect band gap of SnSe. Upon introducing Ta, e.g., Ta$_{0.003}$SnSe, one can find not only the rigid band shift towards the higher binding energy, but also the self-energy broadening of SnSe bands by an impurity scattering, each of which is consistent with the overall energy shift and the fuzzy ARPES intensity maps, respectively, observed in Fig.~2(a-c).

Figure~3 summarizes the energy shift as a function of Ta coverage. The VBM relative to $E_{\rm F}$ was extracted in two different ways, i.\,e.\,, the direct comparison between experimental and calculated results as shown in Fig.~2 and the determination of a cutoff energy of the photoelectron intensity near $E_{\rm F}$. The dashed line at $E_{\rm F}-E_{\rm VBM}=0.45~{\rm eV}$ is half the energy gap~\cite{wang,parenteau} or  $E_{\rm F}$ of undoped SnSe which is a border determining the charge carrier type, e.g., {\it p}- and {\it n}-type for lower and higher value, respectively. The pristine SnSe sample is a {\it p}-type semiconductor with an $E_{\rm F}-E_{\rm VBM}$ of 0.32~eV. As soon as 0.025~ML of Ta is adsorbed on the surface, the shift is 0.19$\pm$0.01~eV, i.\,e.\, $E_{\rm F}-E_{\rm VBM}=0.51\pm0.01~{\rm eV}$, crossing the borderline at $E_{\rm F}-E_{\rm VBM}=0.45~{\rm eV}$ that divides the charge carrier type. With further Ta adsorption, $E_{\rm F}-E_{\rm VBM}$ increases as much as 0.60~eV at 0.15~ML, indicating deeper {\it n}-type semiconducting behavior of SnSe. Based on the change of the electron band structure of SnSe with Ta adsorption, the charge carrier type of SnSe is modified from {\it p}-type, that of the bulk, to {\it n}-type. 

\begin{figure}[t]
\centering
\includegraphics[width=1\columnwidth]{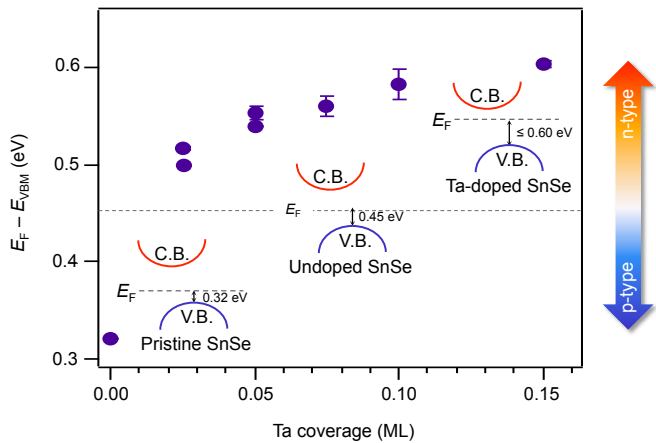}
\caption{\label{fig:fig3} (Color online) The VBM ($E_{\rm VBM}$) relative to $E_{\rm F}$ as a function of Ta coverage. While the energy gap is 0.9~eV~\cite{wang,parenteau}, the criterion determining the carrier type is considered to be 0.45~eV denoted by the dashed line. A pristine sample gives $E_{\rm F}-E_{VBM}=0.32~{\rm eV}$, indicating {\it p}-type carriers. Ta adsorption leads to the shift of the VBM, indicating the change of charge carrier type to {\it n}-type.}
\end{figure}

The origin of the change of the charge carrier type is attributed to the charge transfer from Ta to SnSe. When considering electron affinity, Ta exhibits the lowest electron affinity among Sn, Se, and Ta. In addition, although SnSe is an insulator with an energy gap of 0.9~eV~\cite{wang,parenteau}, it is supposed to exhibit impurity states in the gap region especially when the investigated sample is {\it p}-type semiconductor due to Sn vacancies~\cite{jaekwang}. As a result, electrons are expected to be transferred from Ta to SnSe. This charge transfer results in the shift of the electron band structure, that is exactly manifested in both experimental and theoretical results shown in Fig.~2. The shift is somewhat different from the observed tunable band gap in SnSe induced by K adsorption~\cite{zhang}. Especially, 0.3-2.6~ML of K adsorption leads to electron doping to the conduction band, whereas the valence band moves towards lower binding energy, resulting in the decrease of the band gap. On the other hand, Ta adsorption brings about electron doping to the whole band structure with far less amount of adsorbates, e.\,g.\,, 0.025~ML. Additional interesting feature of the charge transfer is that it is highly localized to the top-most atomic layer of SnSe. A previous theoretical work shows that the major charge density difference, induced by the formation of the interface between Sn and Ta, occurs at the top-most atomic layer of the two atomic-layer unit cell of SnSe~\cite{xian}. When it comes to the fact that ARPES is very sensitive to the top-most layer due to the electron mean free path less than a few {\rm \AA} at 65~eV, used in the experiments, in conjunction with the previous theoretical work~\cite{xian}, the observed energy shifts in Fig.~2(a-c) is considered to be that of the top-most layer of SnSe, whereas bulk SnSe remains intact.

\begin{figure}[t]
\centering
\includegraphics[width=1\columnwidth]{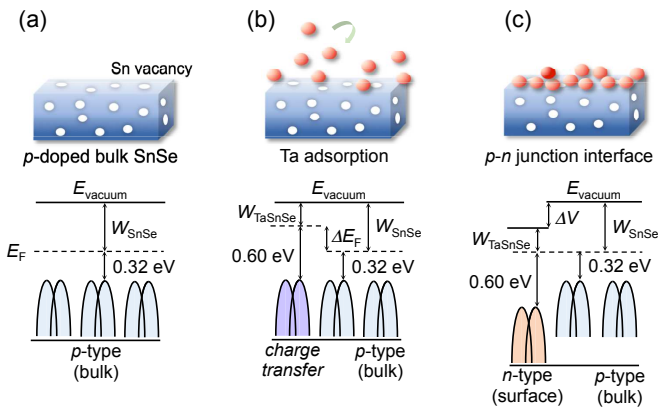}
\caption{\label{fig:fig2} (Color online) (a-c) Upper panels show schematics of the formation of a {\it p-n} junction interface between {\it p}-doped SnSe bulk and Ta-adsorbed SnSe surface. Lower panels show the change of an energy band diagram upon Ta adsorption.}
\end{figure}

Figure~4 shows the schematics of the change of the electron band structure of SnSe induced by Ta adsorption on the surface. Pristine SnSe is a {\it p}-type semiconductor, in which $E_{\rm VBM}$ is at 0.32~eV below $E_{\rm F}$ as shown in panel~(a). $W_{\rm SnSe}$ and $E_{\rm vacuum}$ denote the work-function of SnSe and vacuum energy level, respectively. Figure~4(b) shows a transient state induced by Ta adsorption. Upon introducing Ta, while the charge transfer occurs localized to the top-most layer of SnSe that contacts the adsorbed Ta atoms, $E_{\rm F}$ and hence work-function shift by $\Delta\,E_{\rm F}$ due to the charge transfer. $W_{\rm TaSnSe}$ denotes the work-function of Ta-adsorbed SnSe. In an equilibrium state, when $E_{\rm F}$ of the bulk and the surface of SnSe is aligned as shown in panel~(c), the $E_{\rm VBM}$ of Ta-adsorbed SnSe is observed at higher binding energy, compared to that of the pristine SnSe or the bulk. The resultant potential difference $\Delta\,V$ ($=\Delta E_{\rm F}$), induced at the interface between Ta-adsorbed top-most layer of SnSe and bulk SnSe, can allow electrons free to diffuse from the {\it n}-region to the {\it p}-region~\cite{henck,khomyakov}, leading to the formation of a vertical {\it p-n} junction with an atomically ideal interface and atomically thin {\it n}-region.

In the case of a conventional {\it p-n} junction using bulk materials, the recombination of electron-holes occurs at the junction interface, resulting in a charge depletion layer and band bending. However, in an atomically thin {\it p-n}  junction, the depletion layer is hardly defined due to the reduced thickness~\cite{lee,frisenda}. When a {\it n}-type region is atomically thin, even though the whole bulk is in a {\it p}-type region, the depletion region cannot be formed, so that sharp discontinuity instead of band bending is expected at the interface as schematically shown in Fig.~4(c). In this case, when forward bias is applied, the current is controlled by tunneling-mediated interlayer recombination that overcomes the potential barrier that is described by a combination of both Langevin recombination and Shockley-Read-Hall recombination~\cite{lee,frisenda}. Such an ideal {\it p-n} junction interface will make it possible to fabricate a {\it p-n} junction thermoelectric device, suggested in a previous theoretical study~\cite{fu}. When the internal electric field at the interface is coupled with temperature gradient, it can develop steady current vortices and a magnetic field without external carrier injection. Especially, when the {\it n}-type region is as thin as the depletion region, as is the case of Ta-adsorbed SnSe, the thermopower (or Seebeck coefficient) is predicted to be drastically enhanced within the dynamic electro-thermal model~\cite{Hefner}.

\section{Summary}

The evolution of the electron band structure of SnSe as a function of Ta coverage has been investigated using ARPES technique. When pristine SnSe is a {\it p}-type semiconductor, Ta adsorption of as little as 0.025~ML leads to the shift of the overall band structure as much as 0.2~eV towards the higher binding energy, converting the charge carrier to {\it n}-type. Our theoretical work combined with a previous one shows that charge transfer from Ta to SnSe, localized at the top-most atomic layer, induces the band shift, suggesting the possible formation of an ideal {\it p-n} junction interface with atomically thin {\it n}-region. These findings will shed light on the development of a high efficiency thermoelectric device for future energy application.

\section{Acknowledgements}
This work was supported by the National Research Foundation of Korea (NRF) grant funded by the Korea government (MSIT) (No.~2018R1A2B6004538). The Advanced Light Source is supported by the Office of Basic Energy Sciences of the U.S. Department of Energy under Contract No. DE-AC02-05CH11231. K.K. also acknowledges support by the NRF grant funded by the Korea government (MSIT) (No.~2016R1D1A1B02008461) and Max-Plank POSTECH/KOREA Research Initiative (Grant No. 2016K1A4A4A01922028). H.R. also acknowledges support by the KIST institutional Program (2E29410).

\end{document}